\documentclass[%
 reprint,
superscriptaddress,
 amsmath,amssymb,
 aps,
prl,
]{revtex4-1}

\usepackage{graphicx}
\usepackage[dvipsnames]{xcolor}
\usepackage{amssymb}
\usepackage{amsmath}
\usepackage{amsbsy}
\usepackage{color}

\usepackage{bm}

\usepackage{siunitx}

\begin{document}

\title{Single Molecule Structure of Partially Digested Kinetoplast DNA Networks}
\title{Topological Digestion of Kinetoplast DNA Networks Yield Perturbed Morphologies}
\title{Single-Molecule Morphology of Topologically Digested Olympic Networks}

\author{Saminathan Ramakrishnan}
\affiliation{School of Physics and Astronomy, University of Edinburgh}

\author{Zihao Chen}
\affiliation{School of Biological Sciences, Ashworth laboratories, Institute for
Immunology and Infection Research, University of Edinburgh}

\author{Yair Augusto Gutierrez Fosado}
\affiliation{School of Physics and Astronomy, University of Edinburgh}

\author{Luca Tubiana}
\affiliation{Physics Department, University of Trento}

\author{Willem Vanderlinden}
\affiliation{School of Physics and Astronomy, University of Edinburgh}

\author{Nicholas Jon Savill}
\affiliation{School of Biological Sciences, Ashworth laboratories, Institute for
Immunology and Infection Research, University of Edinburgh}

\author{Achim Schnaufer}
\affiliation{School of Biological Sciences, Ashworth laboratories, Institute for
Immunology and Infection Research, University of Edinburgh}

\author{Davide Michieletto}
\affiliation{School of Physics and Astronomy, University of Edinburgh}
\affiliation{MRC Human Genetics Unit, Institute of Genetics and Cancer, University of Edinburgh}

\newcommand{\dmi}[1]{\textcolor{RoyalBlue}{#1}}

\begin{abstract}
The kinetoplast DNA (kDNA) is the archetype of a two-dimensional Olympic network, composed of thousands of DNA minicircles and found in the mitochondrion of certain parasites. The evolution, replication and self-assembly of this structure are fascinating open questions in biology that can also inform us how to realise synthetic Olympic networks \textit{in vitro}. To obtain a deeper understanding of the structure and assembly of kDNA networks, we sequenced the \textit{Crithidia fasciculata} kDNA genome and performed high-resolution Atomic Force Microscopy (AFM) and analysis of kDNA networks that had been partially digested by selected restriction enzymes. We discovered that these topological perturbations lead to networks with significantly different geometrical features and morphologies with respect to the unperturbed kDNA, and that these changes are strongly dependent on the class of DNA circles targeted by the restriction enzymes. Specifically, cleaving maxicircles leads to a dramatic reduction in network size once adsorbed onto the surface, whilst cleaving both maxicircles and a minor class of minicircles yields non-circular and deformed structures. We argue that our results are a consequence of a precise positioning of the maxicircles at the boundary of the network, and we discuss our findings in the context of kDNA biogenesis, design of artificial Olympic networks and detection of \textit{in vivo} perturbations.
\end{abstract}


\maketitle

The mitochondrial genome of \textit{Kinetoplastid} parasites displays one of the most unique and complex topologies in nature~\cite{Simpson1967,Laurent1970,Shlomai1983,Perez-Morga1993,Shlomai1994,Morris2001,Lukes2002}. The so-called ``kinetoplast DNA'' (i.e. associated with the cellular body, or ``plastos'', near the parasite flagellum that give it its movement, or ``kinetikos'') is a unique genome with a complex topology. In the organism \textit{C. fasciculata}, it is formed by around 5000 interlinked DNA minicircles (2.5 kb), and around 30 larger DNA maxicircles (30 kb). The DNA rings are assembled and replicated into a two-dimensional (2D) network and contained in a membraneless DNA-dense region of $1\mu$m $\times$ $0.4\mu$m within the mitochondrion. The maxicircles mostly encode rRNAs and mRNAs for oxidative phosphorylation and mitoribosomes, while the minicircles encode guide RNA genes required for post-transcriptional editing of the mRNAs~\cite{Simpson1967,Simpson2000,Hajduk2010}.
Kinetoplast DNA replication and biogenesis are not understood and are topics of intense debate in the parasitology community~\cite{Perez-Morga1993,Liu2005,Klingbeil2001a,Hoffmann2018,Schnaufer2010,Kalichava2021,Amodeo2021,Amodeo2022}.

There are several open questions in the field of Trypanosome and \textit{Kinetoplastid} biology; for instance, it is unclear (i) how kDNA was evolutionarily preferred over other simpler forms of genomes (for instance longer DNA rings, as in human mitochondria) and (ii) whether each genetic class of rings (e.g. maxicircles and minicircles) occupy specific and distinct positions within the kDNA structure or whether they are uniformly and randomly dispersed~\cite{Lukes2005}. Interestingly, a strain of \textit{Trypanosome Brucei} can be evolved -- under certain conditions -- to survive and replicate without kDNA~\cite{Schnaufer2002,Schnaufer2010}. Overall, there are a number of unanswered questions about the evolutionary advantage of this structure and whether it is limited to a specific life stage of the parasite. 

Recently, the bio- and polymer physics community used \textit{C. fasciculata} kDNA as archetype of a 2D polymer, which is otherwise challenging to realise synthetically~\cite{Soh2020,Soh2021,Klotz2020,Yadav2023}. In general, the polymer physics of networks of interlinked rings, be it 1D, 2D or 3D, have not been studied thoroughly mainly because of experimental challenges~\cite{Grosberg2020,Tubiana2022}. Thus, it remains rather overlooked how the topology of so-called ``Olympic'' networks affects their material properties, although preliminary works suggest that they may display unique features, such as strong non-linear stress response~\cite{Vilgis1997,Wu2017}, weaker swelling~\cite{Fischer2015,Lang2014a}, percolation~\cite{Igram2016,Diao2012,Michieletto2014} and active elastic tuning~\cite{Kim2013a,Krajina2018}. Linking this back to the biogenesis of kDNA in \textit{Kinetoplastids}, it remains to be determined which -- if any -- of these properties are required for the biological functions of kDNA in the cell. 

Simulations of polymer rings with tunable linking degree suggest that a mean linking degree, or valence, of 3 -- similar to that found in kDNA structures~\cite{Chen1995,Ibrahim2019,He2023} -- may reflect the fact that these networks are poised at the percolation point, where a random graph starts displaying a system-spanning component~\cite{Michieletto2014,Diao2012,Ubertini2023}. In fact, a random graph with valence 3 is optimally connected, i.e. it displays percolation but without redundant links between circles, thus ensuring integrity of the structure whilst optimising the rate of replication, which in \textit{C. fasciculata} occurs through decatenation of minicircles from the network~\cite{Michieletto2014,Perez-Morga1993}. At the same time, whilst the material and elastic properties of kDNA networks are still largely unknown, recent work has estimated the bending stiffness to be in the range of $\kappa \simeq 10^{-21} - 10^{-19} J$~\cite{He2023,Klotz2020} and in-plane Young modulus $Y \simeq 0.1$ pN/$\mu$m~\cite{He2023}. Since the elasticity of an Olympic network correlates with the mean linking number~\cite{Vilgis1997,Ubertini2023,Palombo2023}, we expect that the kDNA elasticity will strongly depend on its underlying topology.

To achieve a better understanding of the connection between material properties and underlying topology, precise single-molecule measurements of experimental Olympic structures are needed. To date, the only synthetic Olympic structures were 1D polycatenanes~\cite{Wu2017,Datta2020,Peil2020}, and bulk 3D Olympic gels~\cite{Kim2013a,Krajina2018} -- the latter made from DNA plasmids and type 2 topoisomerase. Despite these recent advances in the synthesis of catenated structures, there is no experimental method that can precisely quantify the topology of an Olympic network (especially 3D ones). Qualitative single molecule characterisation of kDNA networks have been performed using electron microscopy~\cite{Laurent1970,Barker1980,Ferguson1994} and more recently atomic force microscopy (AFM)~\cite{Cavalcanti2011,Yaffe2021}. 

In our most recent AFM study~\cite{He2023} we performed quantitative image analysis and discovered that AFM, coupled to molecular dynamics simulations, can provide insights into the topology of Olympic networks at single molecule resolution. In this work, we use AFM, quantitative image analysis and polymer theory to understand more about the structure of kDNA and Olympic networks in general. Specifically, by exploiting the fact that catenated DNA circles can be cleaved by sequence-specific  restriction enzymes, we perturb the topology of the network and measure the changes in network morphology. Whilst topological perturbations to kDNA have been explored in the literature (see Refs.~\cite{Chen1995,Yadav2023,Ragotskie2023a}), existing works have not quantitatively measured changes in network morphologies. More specifically, (i) Cozzarelli and co-authors were able to correctly determine the average valence of \emph{C. fasciculata} kDNA by progressively linearising some of its minicircles with XhoI restriction enzyme~\cite{Chen1995}; (ii) Yadav and co-authors measured the shape auto-correlation, and hence the relaxation properties, of networks that had been treated with different restriction enzymes via imaging and discovered that more digested structures had a longer relaxation time, in line with a reduction in network connectivity~\cite{Yadav2023}; (iii) Ragotskie and co-authors caused light-induced nonspecific damage to \textit{C. fasciculata} kDNA and observed the persistence of a 1D edge-loop, consistent with the estimation that the kDNA border is at least 4-fold redundantly linked with respect to the inner minicircles~\cite{Ragotskie2023a}. In our work we wanted to complement these studies and provide a single-molecule view of kDNA networks that underwent specific topological perturbations. 

To achieve \textit{accurate} and \textit{specific} topological perturbations, we first deep sequenced \textit{C. fasciculata} kDNA and discovered that it contains 18 classes of minicircles, the most abundant of these covering 85\% of the network. Through the sequencing, we were able to identify specific restriction endonucleases (REs) that digest different parts of the kDNA structure. We then employed some of these REs to partially digest kDNA and discovered that, for instance, cleaving maxicircles unexpectedly yields significantly shrunk networks. On the contrary, cleaving both the minor class ($\sim$ 10\% of total)  and maxicircles yields significantly larger and non-circular structures. We qualitatively explain these findings with a scaling theory, as kDNA networks with less mass ought to display weaker adsorption to the surface whilst less connected networks are expected to display a weaker bulk modulus and to extend more along the surface. Finally, by cleaving all, and only, minicircles we discovered that most of the maxicircles in the kDNA are interlinked with each other. We argue that our results shed some light on this fascinating and unique structure and potentially inform assembly strategies to synthetically design 2D Olympic networks.

\begin{figure*}[t!]
    \centering
    \includegraphics[width=1\textwidth]{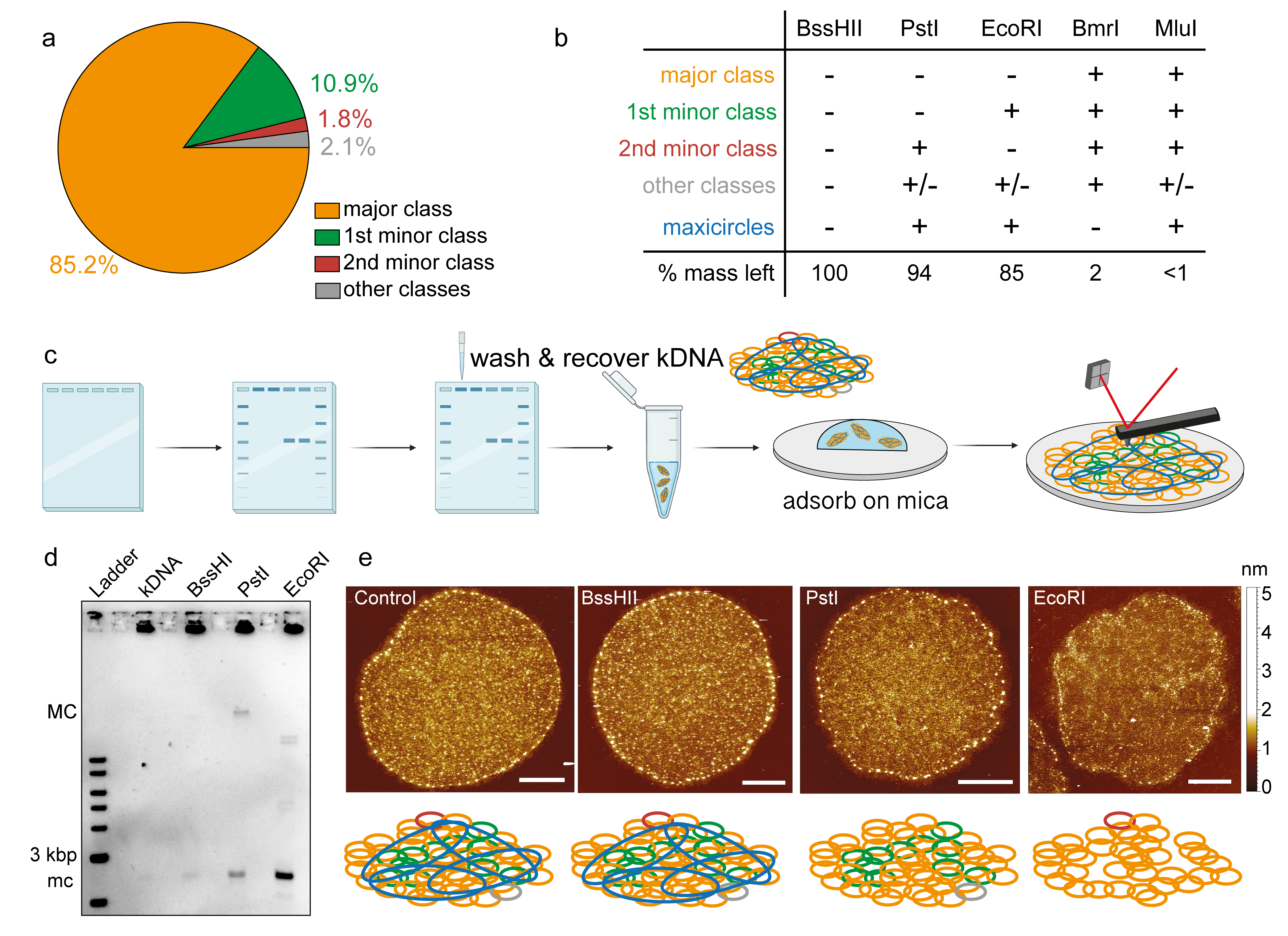}
    \vspace{-0.8 cm}
     \caption{ \textbf{a} Minicircles relative copy number from \textit{C. fasciculata} kDNA sequencing and assembly (see methods). \textbf{b} Table with examples of restriction enzymes that can partially digest (+) or not (-) kDNA components (see SI for a full table with 243 enzymes). $\pm$ denotes digest of some of the ``other'' classes. \textbf{c} Protocol for preparation of clean samples of partially digested kDNA: we load partially digested samples in wells and run gel electrophoresis in a 1.5\% agarose gel at 10V/cm for 20 minutes. The intact and partially digested kDNA structures are too big to travel through the gel and remain stuck in the wells. We flush the wells and recover the partially intact kDNA, which are now cleaned from linearised products and enzymes. \textbf{d} Gel electrophoresis showing partially digested kDNA samples. Linearised maxi circles (MC) run at about 30 kbp, while minicircles (mc) run at 2.5 kbp. \textbf{e} From left to right: images of undigested, BssHII-, PstI- and EcoRI-treated samples. The height color scale is the same throughout the paper. At the bottom, we show colorcoded sketches of the networks: blue, maxicircle; orange, major minicircle class; green, 1st minor minicircle class; red and grey, 2nd and other minicircle classes. } 
     \vspace{-0.4 cm}
    \label{fig:fig1}
\end{figure*}

\section{Results}
\subsection{Deep Sequencing of \textit{C. fasciculata} kDNA reveals 18 classes of minicircles}
Kinetoplast DNA from \textit{C. fasciculata} was purchased from Inspiralis at 100 ng/$\mu$l in TE buffer (10 mM Tris-HCl pH7.5, 1 mM EDTA). A sample of kDNA was sent for deep sequencing at NovoGene. We then performed \textit{de novo} DNA assembly of 4.7 million pair-end reads (150 bp) using KOMICS~\cite{Geerts2021}. We performed several quality checks, including searching for the universally conserved minicircle sequence CSB3 (GGGGTTGGTGT) and comparing our contigs with known sequences of CfC1~\cite{Sugisaki1987}, \textit{T. congolense} and \textit{T. brucei}~\cite{Cooper2019} (see Methods). Overall, the final assembly incorporated over 96\% of all the 4.7 million reads and displayed complete coverage. We detected 18 distinct classes of minicircles with less than 75\% sequence identity, and their relative abundance was estimated from mean read depth calculated by samtools~\cite{Li2009}.  
The major class composes 85.2\% of the kDNA, a first minor class 10.9\%, a second minor class 1.8\% and the other classes make up about 2.1\% of the kDNA (Fig.~\ref{fig:fig1}a). Assuming that the kDNA network contained 5000 minicircles, we estimated 9 to 10 maxicircles per network~\cite{Chen1995}.  

\textit{C. fasciculata} maxicircle genes boundaries were predicted with \textit{Leishmania major} maxicircle annotation to extract unedited genes. We used published transcriptomic data of \textit{C. fasciculata} from \textit{in vitro} culture and mosquito hindguts to validate unedited maxicircle gene annotations and predict edited encrypted genes~\cite{Filosa2019}. T-masked mapping using T-aligner~\cite{Gerasimov2018} suggested that CfC1 capable of editing four cryptic genes: ATPase subunit 6 (A6), ribosomal protein S12 (RPS12 or uS12m), NADH dehydrogenase subunit 7 (ND7), and cytochrome oxidase subunit 2 (COXII). Guide RNAs were identified on 13 out of 18 minicircle classes, which captured the same gRNAs on contigs containing the annotated minicircle fragments in previous study~\cite{Gerasimov2018}. However, no gRNA was found on the major and the 1st minor minicircle classes.

Finally, RE cutting sites on \textit{C. fasciculata} minicircles and maxicircle were predicted for enzymes available from New England Biolabs (see Fig.~\ref{fig:fig1}b and SI for a full table). Specifically, we chose BssHII as negative control, PstI as cutting maxicircles and a small $< 4\%$ fraction of minicircles, EcoRI cutting maxicircles and around 13\% of the minicircles and finally BmrI, cutting all, and only, minicircles. Through these enzymes, we specifically implement topological perturbations to the kDNA structure with the aim of quantifying specific changes in their morphology via Atomic Force Microscopy.

\begin{figure*}[t!]
    \includegraphics[width=1\textwidth]{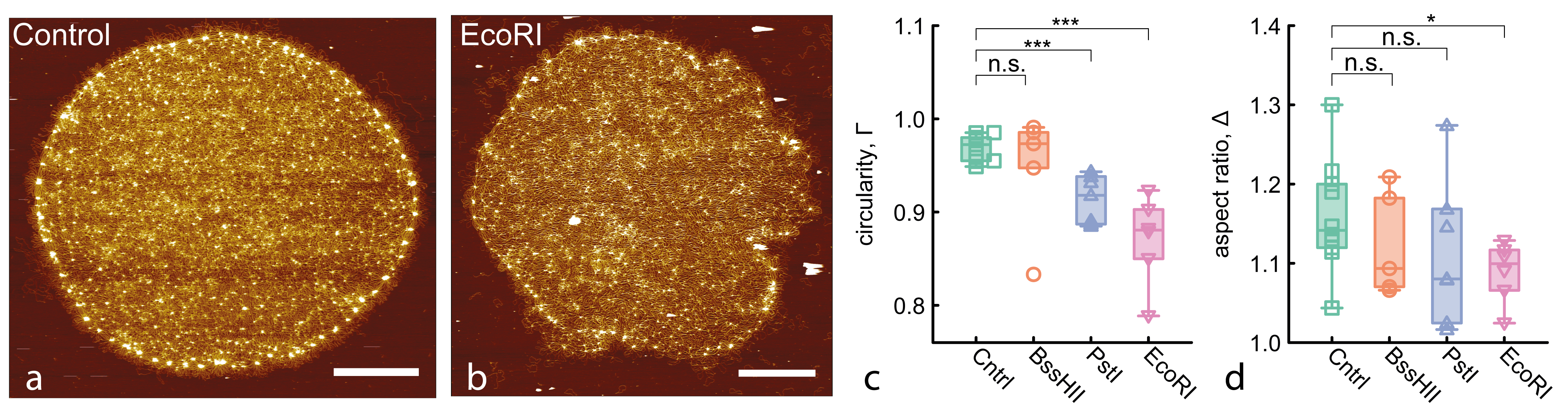}
    \vspace{-0.8cm}
    \caption{ \textbf{a-b} Example AFM images of control and EcoRI treated samples. Scale bar is 2 $\mu$m. \textbf{c} Boxplot showing the circularity $\Gamma$. Smaller $\Gamma = 4 \pi (\textrm{area}/\textrm{perimeter}^2)$ indicate that the perimeter is longer than the one of a perfect circle and it quantifies irregular kDNA borders. \textbf{d} Boxplot showing aspect ratio $\Delta = M/m$ between major and minor axes. Lower aspect ratios indicate shapes with more similar major and minor axes. EcoRI digested structures display more irregular borders but the overall shape is less ellipsoidal and more circular. }
    \label{fig:fig_shape}
\end{figure*}

\subsection{Preparation of partially digested kDNA for AFM}

For restriction digestion with EcoRI, PstI and BssHII, 1 $\mu$L of enzyme (10 Units) was used to digest 1 $\mu$g of kDNA in 1X rCutsmart buffer, overnight at 37$^\circ$C. The BssHII-treated sample was also incubated at 50$^\circ$C for 2 hours prior to overnight incubation at 37$^\circ$C, as per NEB recommendation. We found that during the AFM sample preparation, the mica surface was quickly covered by recombinant albumin in rCutSmart buffer and restriction digested mini and maxicircles, particularly at high magnesium concentration. Thus, enzyme-treated partially digested kDNA structures were poorly adsorbed on the mica surface. To solve this issue, we developed a methodology to remove linearised mini and maxicircles, enzymes and albumin for AFM sample preparation as follows (see also Fig.~\ref{fig:fig1}c): the kDNA sample was first prestained with diluted SybrGold and ran in 1.5\% agarose gel at about 10 V/cm for 20 minutes (see Fig.~\ref{fig:fig1}c,d). The gel tray was then removed from the tank and placed on a UV transilluminator. 1X TE buffer in the well was gently pipetted out and replaced with 80 $\mu$L of adsorption buffer (10 mM Tris-HCl pH 7.9, 10 mM MgCl$_2$, 1 mM EDTA). This step was repeated twice and 50 $\mu$L of adsorption buffer was left in the well after the second wash. Under UV light, a fluorescent layer was visible on the wall of the well; this is because the kDNA structure contains thousands of DNA rings that can individually travel through the gel, but remain stuck in the well when linked with each other in the (partially) intact kDNA (Fig.~\ref{fig:fig1}d). More specifically, we found that the kDNA becomes weakly adsorbed on the wall of the well and is easily removable by washing. The kDNA was either gently flushed using a pipette with 10-20 $\mu$L of adsorption buffer or gently touched with a pipette tip to resuspend the kDNA back into the buffer. We took utmost care to avoid disrupting kDNA integrity at this stage. After resuspension, a 50 $\mu$L of sample was recovered and adsorbed on freshly cleaved mica for 10 minutes. 
The sample was dip washed in ultra-pure water for 1 minute and gently air-dried in ultrapure nitrogen stream (see Fig.~\ref{fig:fig1}e and SI for example images). 

\begin{figure*}[t!]
    \includegraphics[width=1\textwidth]{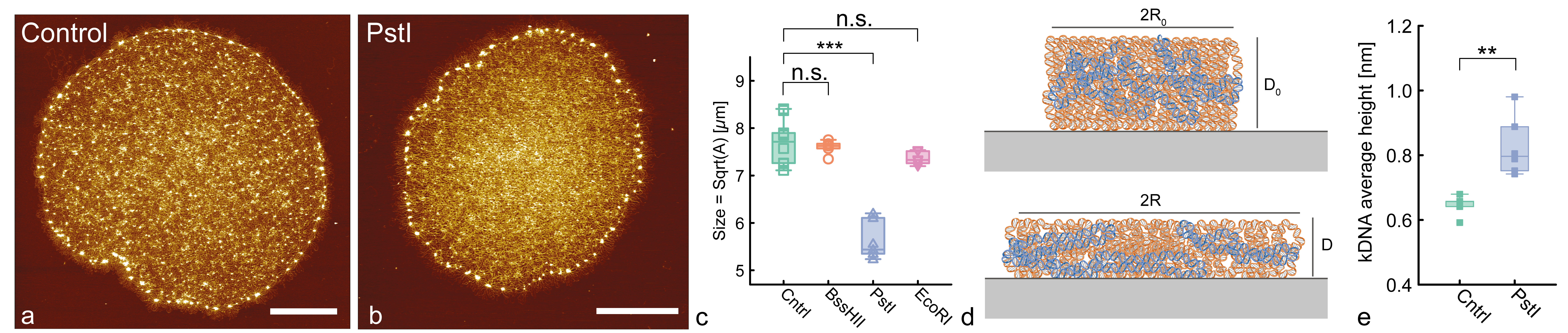}
    \vspace{-0.3cm}
    \caption{ \textbf{a-b} Examples of AFM images of control and PstI treated samples (scale bar is 2 $\mu$m). \textbf{c} Boxplot of the diameter $d = \sqrt{\textrm{area}}$ of kDNAs. \textbf{d} Schematics sketching weak (top) and strong (bottom) adsorption of kDNA to the mica. \textbf{e} Boxplot of the average median height of the kDNA structures for control and PstI-treated samples (see SI for details).}
    \label{fig:fig_size}
\end{figure*}

\subsection{Partially digested structures display irregular borders}


First, we noticed that EcoRI-treated networks (missing all maxicircles and around 13\% of minicircles) appeared structurally disrupted, while the others were, in first approximation, largely unperturbed (Fig.~\ref{fig:fig1}e). To quantify this change in morphology we measured the circularity $\Gamma$ and aspect ratio $\Delta$ (see Fig.~\ref{fig:fig_shape}). To obtain these two quantities, we manually traced the closed contour of the kDNA structures in ImageJ, and computed the circularity as $\Gamma =  4 \pi (A/p^2)$, where $A$ is the area of the closed curve and $p$ its perimeter, while the aspect ratio as $\Delta = M/m$ where $M$ and $m$ are the major and minor axes of a fitted ellipse. A value of $\Gamma = 1$ indicates that the perimeter is that of a perfect circle, whilst $\Gamma < 1$ indicates longer, typically irregular, perimeters. At the same time, $\Delta$ refers to the shape of the overall object, with $\Delta = 1$ indicating an object with near circular symmetry, while $\Delta \simeq 0$ indicating a rod. 

As shown in Fig.~\ref{fig:fig_shape}c, we observe a significant (p-value $< 0.001$) decrease in circularity in both PstI and EcoRI treated samples, but no significant change of $\Gamma$ in our BssHII control. The value of $\Gamma$ drops from near 1 for the control to below 0.9 for EcoRI treated samples, reflecting the appearance of irregular kDNA borders, almost ``blebbing'', visually evident in the AFM images (Fig.~\ref{fig:fig_shape}a-b). Additionally, we find a decrease, albeit more modest, in the aspect ratio $\Delta$ reflecting more symmetric structures, with a smaller difference between major and minor axes (Fig.~\ref{fig:fig_shape}d). This confirms that the control samples are somewhat ellipsoidal with a well defined major axis roughly 20\% longer than the minor axis. 

Interestingly, the fact that PstI-treated samples show a minor change in circularity $\Gamma$ suggest that they contribute to the structural integrity of the network, especially of the border. It is also interesting to note at this stage that the mass contribution of maxicircles is about 2.3\%, i.e. around 200 $10^3$ kg/mol within a total of 8.5 $10^6$ kg/mol for the whole kDNA. Fig.~\ref{fig:fig_shape}c then suggests that cleaving both maxicircles and the minor class of minicircles (in total about 15\% of the kDNA mass) has a compound effect in the network morphology likely due to a decrease in network connectivity~\cite{Chen1995,Yadav2023}.

\subsection{Cleaving maxicircles yields significantly shrunk networks}

To understand the specific structural role of maxicircles, we decided to focus on PstI-treated kDNA structures, which lack maxicircles and a small ($< 4\%$) fraction of minicircles. First, we validated in our agarose gel that a small number of minicircles are  cleaved along with maxicircles in PstI-treated samples (Fig.~\ref{fig:fig1}d). Second, and more importantly, we observe that PstI treatment causes a dramatic reduction in kDNA area $A$, or diameter $d = \sqrt{A}$, from around $d = 8$ $\mu$m to around $d = 5.5$ $\mu$m, i.e. a 1.5-fold shrinkage with respect to the control and BssHII-treated samples (Fig.~\ref{fig:fig_size}).  Curiously, and unexpectedly, EcoRI-treated samples appear to recover a larger size compared with PstI-treated kDNAs (Fig.~\ref{fig:fig_size}c). 

We can try to rationalise this puzzling observation with a simple scaling argument: the free energy of adsorption of a soft polymeric cylinder on a surface can be written as~\cite{Gennes1979a} 
\begin{equation}
    \mathcal{F} = k_B T \kappa \dfrac{D_0^2}{D^2} - k_B T \delta f_b N
    \label{eq:freeen}
\end{equation}
where $D$ is the average extension of the polymers away from the surface, $\kappa$ is an effective stiffness that is proportional to the in-plane Young modulus, $\delta$ is an effective interaction between the monomers and the surface and $N$ is the total length of the polymers in the cylinder (see Fig.~\ref{fig:fig_size}d for a schematics). Assuming that the whole kDNA mass is confined within a layer $D$ from the surface, the fraction of mass adsorbed is approximately $f_b \simeq a/D$, where $a$ is the extent of the attractive layer from the surface. We can substitute this into Eq.~\eqref{eq:freeen}, 
\begin{equation}
\mathcal{F} = k_B T \kappa \dfrac{D_0^2}{D^2} - k_B T \delta \dfrac{a}{D} N
\end{equation}
and minimise it with respect to $D$, i.e. $\partial \mathcal{F}/\partial D = 0$, to obtain 
\begin{equation}
\dfrac{D}{D_0} = 2 \dfrac{D_0}{a} \dfrac{\kappa}{\delta N} \sim \kappa (\delta N)^{-1} \, .
\end{equation}
This equation implies that the average height of kDNA away from the adsorbing surface is inversely proportional to the effective adsorption strength $\delta N$, and directly proportional to the stiffness $\kappa$, respectively. Assuming that the whole kDNA is contained within a constant cylindrical volume $V = \pi R_0^2 D_0 = \pi R^2 D$ we obtain a relationship between the average kDNA extension $R$ and the total adsorption energy $\delta N$ scaling as 
\begin{equation}
\dfrac{R}{R_0} \sim \left(\dfrac{\delta N}{\kappa}\right)^{1/2} \, ,
\label{eq:extension}
\end{equation}
implying that the more the kDNA mass, i.e. the larger $N$, the stronger the adsorption strength and the larger the planar extension $R$ of the kDNA network. This scaling argument entails that the lack of maxicircles decreases the overall mass of kDNA, in turn decreasing the net adsorption strength of the kDNA structure. At the same time, as expected, the larger the kDNA in-plane stiffness $\kappa$ the smaller the lateral extension or spreading along the surface.  

This simple scaling argument is in line with our observations: removing maxicircles, i.e. decreasing $\delta N$, reduces $R$, as seen in Fig.~\ref{fig:fig_size}c and increases the average height, as seen in Fig.~\ref{fig:fig_size}e, where we compute the median kDNA height from the mica (see SI for details). On the other hand, we note that the mass lost due to PstI treatment amounts to about 2\% of the total mass (see above). Because of this, we expect a relatively small change in kDNA extension, in marked contrast with the significant (1.5-fold) reduction in kDNA diameter seen in experiments (Fig.~\ref{fig:fig_size}c).  Additionally, according to Eq.~\eqref{eq:extension}, we should expect a similar shrinking in networks where the minor class of minicircles, accounting for about 10\% of kDNA mass, has been cleaved. On the contrary, we did not observe such shrinkage in EcoRI-treated networks, which in fact displayed more extended structures (Fig.~\ref{fig:fig_size}c). These contrasting results may be reconciled by arguing that while the removal of maxicircles leads to a shrinkage of the network due to smaller adsorption strength ($\delta N$), the additional cleavage of a considerable $\sim 10\%$ fraction of minicircles reduces the topological connectivity and the in-plane stiffness ($\kappa$) of the network, in turn allowing easier spreading on the surface. 

\begin{figure}[t!]
    \includegraphics[width=0.45\textwidth]{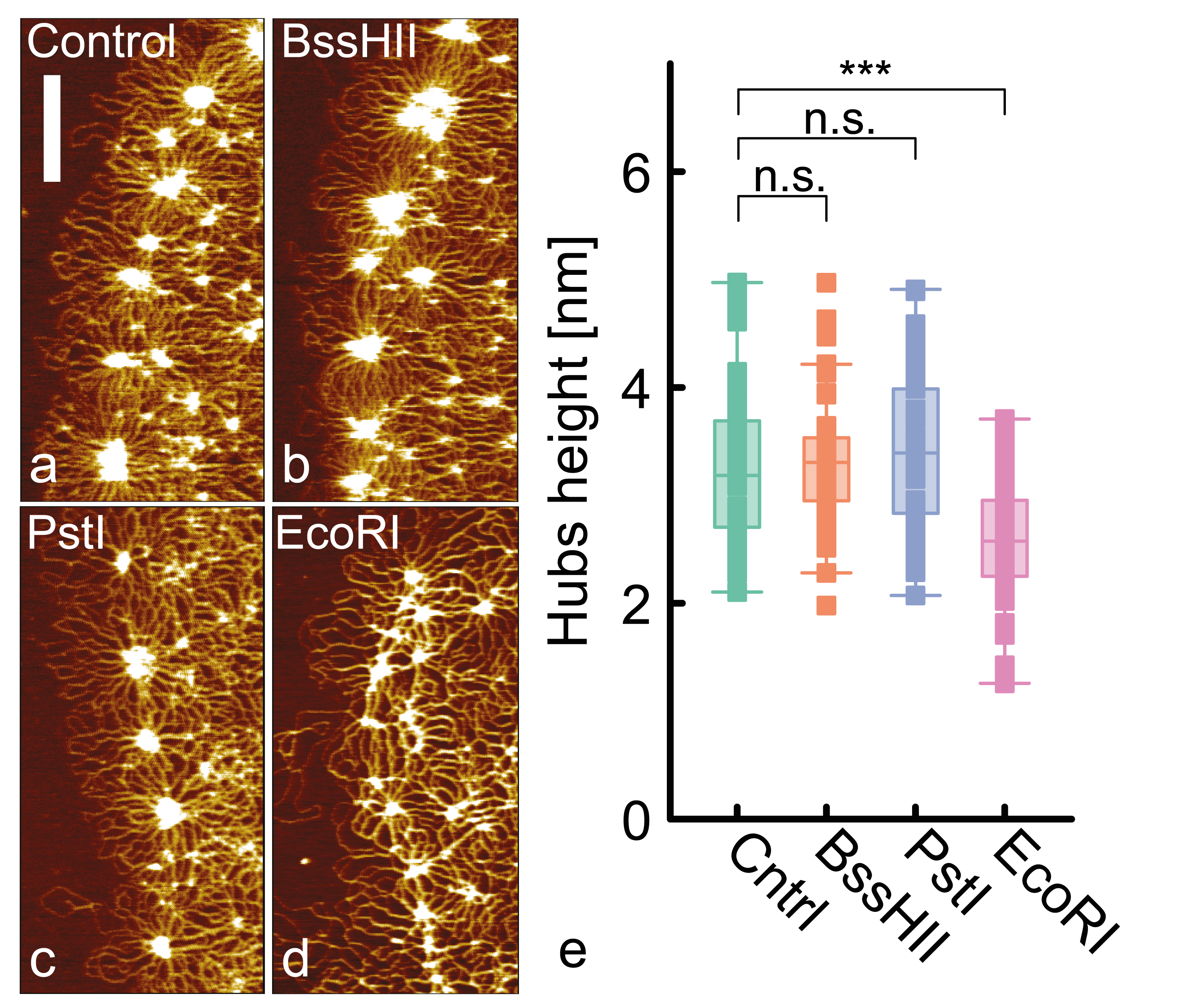}
    \vspace{-0.3cm}
    \caption{\textbf{a-d} Example AFM images (zoomed in at the borders) of control, BssHII, PstI and EcoRI samples. The scale bar is 500 nm. \textbf{e} Boxplot showing the height of the hubs for the different samples. }
    \label{fig:hubs}
\end{figure}

\subsection{EcoRI-treated networks display significantly disrupted hubs}

After having quantified global changes to kDNA morphology, we then turned our focus to smaller scale sub-structures. More specifically, from the AFM images we realised that the hubs, or rosette, structures that characterise intact kDNA are affected by partial digestion. In Fig.~\ref{fig:hubs}a-d we show representative zoomed in sections of kDNAs, showing a series of hubs at the periphery of partially digested kDNAs. To quantify their state, we extract the height of the tallest pixel across several tens of hubs around each kDNA: while BssHII and PstI treated hubs are not significantly affected, EcoRI treatment causes a significant disruption, with the hubs' height reducing from $3.3 \pm 0.7$ nm to $2.6 \pm 0.5$ nm (p-value $< 0.001$, Fig.~\ref{fig:hubs}e). We note that even for the control case, the average height is smaller than the one expected for several (possibly tens) of DNA strands overlapping each other at the hub~\cite{Ragotskie2023} -- each around $2$ nm thick in the hydrated DNA structure -- due to (i) imaging dehydrated DNA and (ii) the compression of the DNA by the AFM tip~\cite{He2023}. Nevertheless, the qualitative and quantitative disruption in the EcoRI treated samples is clear and points to the fact that hubs are mostly made by essential crossings between minicircles rather than maxicircles. At the same time, we note that PstI treated samples do not display significantly disrupted hubs, but do display fewer connections between the hubs (Fig.~\ref{fig:hubs}c) and also more irregular borders (Fig.~\ref{fig:fig_shape}c).

\subsection{Maxicircles form an Olympic sub-network and weave through hubs at the periphery}

\begin{figure*}[t!]
    \includegraphics[width=0.92\textwidth]{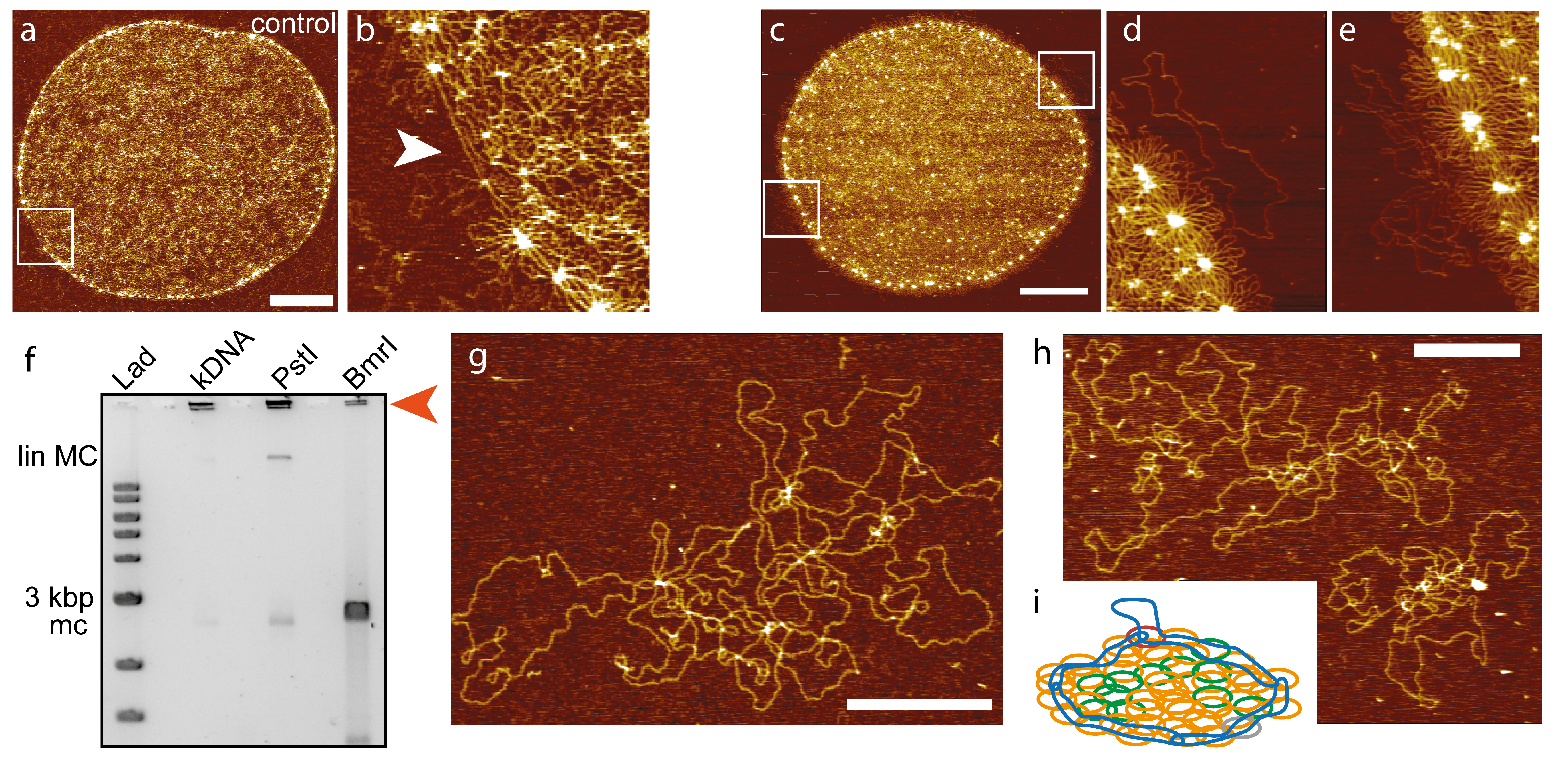}
    \vspace{-0.3cm}
    \caption{ \textbf{a-e} AFM images and zoomed in regions displaying examples of maxicircles weaved at the periphery (white arrows) and threading through minicircles and hubs (white circle). Scale bar in \textbf{a} and \textbf{c} is 2 $\mu$m. \textbf{f} Gel elctrophoresis comparing kDNA, PstI-treated kDNA and BmrI-treated kDNA. The latter enzyme cuts all minicircles but leaves maxicircles intact. The presence of DNA mass in the well (orange arrow) suggests that the maxicircles are linked together. (Lin MC = linearised maxicircles, mc = minicircles). \textbf{g-h} AFM images of maxicircles purified from BmrI-treated samples and displaying linking. Scale bar is 500 nm. \textbf{i} Sketch of kDNA model with some of the maxicircles forming a sub-linked network and weaving around the periphery. }
    \label{fig:maxicircles}
\end{figure*}

Our findings suggest that (i) hubs are mostly made by links between minicircles (Fig.~\ref{fig:hubs}) and (ii) maxicircles contribute to provide structural integrity to the kDNA border (Fig.~\ref{fig:fig_shape}). While it is well known that during kDNA replication the minicircles are polymerized and reattached at the periphery of the parent kDNA by enzymes~\cite{Perez-Morga1993}, the fate of maxicircles and their position within the kDNA after cell division are unknown. Still, we note that during cell division, the maxicircles form the so-called ``nabelschnur'', a DNA bridge connecting newly replicated kDNAs in the daughter cells~\cite{Gluenz2011}, suggesting that they may assume a more peripheral position compared to minicircles. In light of this, and motivated by the dramatic structural and morphological change in PstI treated samples, we hypothesised that at least some maxicircles may be weaved along the kDNA periphery and provide direct support to the border. To qualitatively test this hypothesis, we first visually inspected AFM images of non-digested samples, and observed clear signatures of maxicircles joining hubs and threading minicircles along the periphery of intact kDNAs (see white arrow in Fig.~\ref{fig:maxicircles}a,b) and we also identified cases in which maxicircles were clearly linked near the border of the network and spreading outside it (Fig.~\ref{fig:maxicircles}c-e). To further understand whether the maxicircles are themselves forming a sub-network, we treated kDNAs with BmrI, a restriction enzyme that cuts all minicircles but leaves maxicircles intact. Intriguingly, BmrI-treated samples displayed a consistent DNA band that did not travel into the gel (Fig.~\ref{fig:maxicircles}f). This suggests that maxicircles in \textit{C. fasciculata} are linked with each other, as found in \textit{Trypanosoma equiperdum}~\cite{Shapiro1993}. To further prove this finding, we extracted the DNA mass in the well as before and visualised it under AFM; indeed, we could observe catenated structures of several (but likely not all 10) maxicircles (Fig.~\ref{fig:maxicircles}g-h). Thus, in light of these results we argue that at least some maxicircles are thread and weave along the periphery of the kDNA providing structural integrity and that they form a percolating sub-network within the kDNA.

\section{Discussion}
The kinetoplast DNA remains one of the most mysterious and fascinating structures in nature. Its biogenesis, self-assembly and replication are puzzling and still not fully understood. To address some of the open questions in this field we performed atomic force microscopy (AFM) on \textit{C. fasciculata} kDNA samples that had been partially digested by restriction enzymes. We specifically identified restriction enzymes that cut different fractions of the kDNA structure via deep sequencing and DNA assembly. More specifically, we chose BssHII having no targets, PstI targeting maxicircles and $< 4\%$ of minicircles (in total around $6\%$ of kDNA mass), EcoRI targeting maxicircles and around $13\%$ of minicircles (in total around $15\%$ of kDNA mass) and BmrI cutting all, and only, minicircles in \textit{C. fasciculata} kDNA (Fig.~\ref{fig:fig1}a-b, d). 

First, we proposed a new method to obtain kDNAs that are suitable for AFM: it employs gel electrophoresis to clean samples from enzymes, BSA and other small kDNA fragments. We argue that it could be used in the future as a simple way to purify kDNA and other samples containing catenated structures such as some DNA origami for AFM (Fig.~\ref{fig:fig1}c). We then measured the change in shape and size of these structures and observed that both EcoRI and PstI treated structures displayed a significant reduction in circularity, with the border of PstI and EcoRI treated samples appearing more irregular and ``blebbing'' than control samples. These results suggested that both maxicircles and minicircles play an important role in kDNA border integrity (Fig.~\ref{fig:fig_shape}).

We then reported a marked change in size of PstI treated kDNA samples, with a diameter reduced 1.5-fold when adsorbed on mica with respect to the control sample (Fig.~\ref{fig:fig_size}). Intriguingly, this size change was not observed in confocal microscopy in the bulk~\cite{Yadav2023}. On the contrary, we did not observe any significant shrinking in EcoRI treated samples. We rationalised this finding via a simple scaling argument (see Eq.~\eqref{eq:extension}), which predicts that less polymer mass in the kDNA network reduces the effective adsorption strength but that cleaving minicircles may decrease the network stiffness. These two parameters, kDNA mass and network stiffness, balance each other and both affect the average network extension in opposite ways.

We note that similar kDNA structures have been recently prepared and analysed via confocal microscopy in a bulk solution by Yadav et al.~\cite{Yadav2023}. They reported that the digested kDNAs did not appear to assume a different size or shape, yet they displayed a different dynamical relaxation timescale, which they measured by computing correlations of the anisotropy vector. The increasing internal relaxation timescale was attributed to a smaller internal kDNA connectivity, which rendered the networks floppier and hence slower to relax. In contrast with their findings, we do instead observe a dramatic change in size (1.5-fold reduction) of PstI-treated samples (lacking mainly maxicircles) and a significant reduction of circularity in PstI and EcoRI treated samples (lacking both maxicircles and 13\% of minicircles). Yet, in line with their findings, our results suggest that the EcoRI-treated networks are floppier than the control ones and can stretch more on the mica.

Further, we found that the hubs' height is significantly reduced after EcoRI treatment, suggesting that minicircles are the main component of those structures. At the same time, we found a significant change in border shape after PstI. We thus argue that maxicircles may assume a specific spatial distribution within the network and find visual, qualitative evidence that at least some of them may thread and weave along the border, contributing to the structural scaffolding of the hubs (Fig.~\ref{fig:maxicircles}a-e).  Finally, by treating kDNAs with BmrI we found evidence that maxicircles form a percolated interlinked sub-network (Fig.~\ref{fig:maxicircles}f).

Overall, our single molecule AFM quantitative characterisation of partially digested kDNA offer some insight into kDNA's unique structure. Despite this, more work is needed to exactly pinpoint the spatial distribution of maxicircles and different classes of minicircles, and to dissect how each component contributes to the kDNA material properties. In the future, we plan to analyse and quantitatively compare the kDNA from different parasites; for instance, there are forms of trypanosomes that no longer depend on kDNA function and have lost maxicircles but not minicircles; thus, it would be natural to look at kDNA structures extracted from these strains~\cite{Schnaufer2002}.
We hope that ultimately, our results will complement others, to achieve a full understanding of the biogenesis and self-assembly of this fascinating and mysterious structure.

\section*{Acknowledgements}
DM thanks the Royal Society for support through a University Research Fellowship. This project has received funding from the European Research Council (ERC) under the European Union's Horizon 2020 research and innovation program (grant agreement No 947918, TAP). The authors also acknowledge the contribution of the COST Action Eutopia, CA17139. The minicircle assembly is deposited on GenBank OR687467-OR687484.

\section{Methods}

\subsection*{kDNA sequencing and bioinformatic analysis}

\subsubsection{DNA Assembly}

Kinetoplast DNA from \textit{C. fasciculata} was purchased from Inspiralis at 100 ng/$\mu$l in TE buffer (10 mM Tris-HCl pH7.5, 1 mM EDTA). 

kDNA sample was sequenced at NovoGene. After Microbial Whole Genome Library Preparation (350bp), libraries were pooled and sequenced on Illumina sequencer. Pair-end reads (150 bp) were generated and about 4.7 million reads passed the quality filter.

To optimize the minicircle detection and construction, de novo assembly was performed in KOMICS with increase kmer sizes~\cite{VandenBroeck2020}. Contigs that contained the CSB3 (GGGGTTGGTGT) or its reverse complement, with one allowed mismatch to capture sequence variations, were tested for circularity. Fragments of the published minicircle sequence was detected and was annealed manually for circularization and comparison with the known sequence of major minicircle type in CfC1~\cite{Sugisaki1987}. The circularized contigs were orientated to the same strand and aligned at their anchor region. The published annotated minicircle fragments were detected in the de novo assembled complete minicircles~\cite{Yasuhira1995}. After assembly with small kmers, contigs homologous to \textit{L. braziliensis} maxicircle were extracted for manual examination by global mapping that allowed one mismatch, to select for the longest homologous contig with complete read coverage.


\subsubsection{Relative abundance estimation}

After minicircles of three \textit{T. congolense} strains were assembled, the conserved regions was identified by visually examining aligned minicircles. Motifs homologous to \textit{T. brucei} CSB1, 2, and 3 were recognized in both conserved area~\cite{Cooper2019}. Illumina reads were mapped to the assemblies using Bowtie 2 with \verb|--very-sensitive| option~\cite{Langmead2012a}. Subsequent IGV visualization revealed no region with low or no coverage~\cite{Robinson2011}. The minicircle assemblies incorporated over 96\% of kDNA reads. The completeness was confirmed by $>$98\% mapped CSB3-containing reads and $>$97\% CSB3-containing reads mapped with quality $\geq 10$. More specifically, we obtained 4701610 total reads, mapped 4516344 (96\%) and a total of 494007 CSB3-containing reads, and mapped 489034 (98\%). The relative abundance of 18 unique minicircles was estimated from mean read depth calculated by samtools~\cite{Li2009}. 

\subsubsection{Minicircle Annotation}

\textit{C. fasciculata} maxicircle was aligned to annotated \textit{L. major} maxicircle and subsequently to unedited \textit{L. major} maxicircle genes for annotation. The annotation was confirmed by aligning transcriptomic data from \textit{in vitro} adherent and swimming form of \textit{C. fasciculata} and infected mosquito hindguts~\cite{Filosa2019}. Using the unedited \textit{C. fasciculata} maxicircle genes as a reference, T-aligner performed T-masked mapping to detect consistently edited encrypted genes that were used in subsequent edited gene prediction and minicircle annotation~\cite{Gerasimov2018}. The published A6 and uS12m edited mRNAs~\cite{Yasuhira1995} as well as predicted edited ND7 and COXII mRNAs ~\cite{Filosa2019} were validated by alignment of the transcriptomic data with Bowtie2~\cite{Langmead2012a}. Guide RNA prediction was achieved using a custom version of the kDNA annotation package~\cite{Cooper2019}. Restriction enzyme cutting sites on \textit{C. fasciculata} minicircles and maxicircle were predicted for enzymes available from New England Biolabs (NEB) using a custom python script to facilitate enzyme choice. 

\subsection*{kDNA digestion}
For restriction digestion with EcoRI, PstI, BssHII (all from NEB), 1$\mu$L of enzyme (10 Units) was used to digest 1 $\mu$g of kDNA in 1X rCutsmart buffer, overnight at 37$^\circ$C. The BssHII restriction digestion sample was incubated  at 50$^\circ$C for 2 hours prior to overnight incubation at 37$^\circ$C. The samples were then ran on a gel and recovered from the wells, as described in the text. After recovery, the samples were adsorbed on freshly cleaved mica, dip washed in ultra-pure water for 1 minute and gently air-dried in ultrapure nitrogen stream. 
  
\subsection*{AFM images acquisition and analysis}
The AFM images were recorded in Bruker JPK NanoWizard 4XP using SNL-10 probes. To maintain uniformity in comparing the structural changes, we traced only the circular kDNA structures and recorded the topographs at high resolution (2000 pixels X 2000 pixels). The AFM topographs were post-processed in the JPK data processing software and converted into TIF files. We then used ImageJ to manually compute the area of each kDNA, distance between hubs and pore size using morphological segmentation via MorphoLibJ~\cite{Legland2016} as previously described (Ref.~\cite{He2023}). The heights of the hubs in Figure 4e were quantified using Gwyddion software. For each AFM image, individual height profiles were generated for a minimum of 50 hubs, and the peak maxima for each hub were recorded to calculate the average height in nm.

To obtain the average height of the whole kDNA, we performed an analysis using MountainsSPIP software. First, the AFM images were pre-processed using the least square plane (LSPL) subtraction function to flatten any long-range slope in the AFM images. Subsequently, a line-by-line levelling process was applied by subtracting the least square polynomial and mean values to ensure straightness and align each scan line onto a uniform flat surface within the image. The resulting flat images and data points were then used to generate a histogram of the pixel height distribution (see SI). Within the histogram, two prominent peaks corresponding to the mica surface and kDNA structures (adsorbed DNA) were identified. We then calculated the average height of the kDNA structures by taking the difference between the two maxima of the two peaks. 

\subsection*{Agarose gel electrophoresis}
We prepared 0.5 $\mu$g of kDNA with 0.5$\mu$L of restriction enzymes (BssHII, Pst1, EcoRI) in 1X rCutsmart buffer and incubated overnight at 37$^\circ$C. The BssHII sample was incubated at 50$^\circ$C for 2 hours before the overnight incubation, as per NEB recommendations. A 0.8\% agarose gel was cast in 1X TAE buffer and 25 $\mu$L of restriction digested kDNA samples (plus a control sample) were prepared in 1X loading dye and run for 5 hours at 50V (about 5V/cm). The gel was stained with SYBR gold and imaged in a geldoc imaging system.

\bibliographystyle{apsrev4-1}
\bibliography{library1,library}
\end{document}